\documentclass[11pt]{article}
\usepackage{moriond,epsfig}

\def\NIM{{\em Nucl. Instrum. Methods}}

\def\PLB{{\em Phys. Lett.}  B}
\def\PRL{{\em Phys. Rev. Lett.}}
\def\PRD{{\em Phys. Rev.} D}

\newcommand{\Bz}{B^0}
\newcommand{\Bzb}{\bar B^0}
\newcommand{\Bs}{B^0_s}
\newcommand{\Bsb}{\bar B^0_s}
\newcommand{\barB}{{\bar B}}
\newcommand{\beqn}{\begin{eqnarray}}
\newcommand{\eeqn}{\end{eqnarray}}
\newcommand\Abar{\kern 0.18em\overline{\kern -0.18em A}{}}
\newcommand\Apz{A^{+0}_{\rho\pi}}

\newcommand\Apzb{\Abar^{+0}_{\rho\pi}}
\newcommand\Azp{A^{0+}_{\rho\pi}}
\newcommand\Azpb{\Abar^{0+}_{\rho\pi}}
\newcommand\Azz{A^{00}_{\rho\pi}}
\newcommand\Azzb{\Abar^{00}_{\rho\pi}}
\newcommand\Apm{A^{+-}_{\rho\pi}}
\newcommand\Amp{A^{-+}_{\rho\pi}}
\newcommand\Apmb{\Abar^{+-}_{\rho\pi}}
\newcommand\Ampb{\Abar^{-+}_{\rho\pi}}
\newcommand\Apipi{A^{+-}_{hh}}
\newcommand\Abarpipi{\overline\Apipi}

\newcommand\Apippiz{A^{+0}_{hh}}
\newcommand\Abarpippiz{\overline\Apippiz}

\newcommand\Apizpiz{A^{00}_{hh}}
\newcommand\Abarpizpiz{\overline\Apizpiz}
\newcommand\BRpipi{{\cal B}^{+-}}
\newcommand\BRpippiz{{\cal B}^{+0}}
\newcommand\BRpizpiz{{\cal B}^{00}}

\def\ifmath#1{\relax\ifmmode#1\else$#1$\fi}

\def\bc{\begin{center}}
\def\ec{\end{center}}
\def\bi{\begin{itemize}}
\def\ei{\end{itemize}}
\def\bv{\begin{verbatim}}
\def\ev{\end{verbatim}}

\newcommand{\beq}{\begin{equation}}
\newcommand{\eeq}{\end{equation}} 
\newcommand{\beqa}{\begin{eqnarray}}
\newcommand{\eeqa}{\end{eqnarray}}
\def\CP     {\ifmath{C\!P}}
\def\babar{\mbox{\sl B\hspace{-0.4em} {\scriptsize\sl A}\hspace{-0.4em} B\hspace{-0.4em} {\scriptsize\sl A\hspace{-0.1em}R}}}
\newcommand{\comment}[1]{}

\def\Kbar{\ifmath{\kern 0.2em\overline{\kern -0.2em K}}{}}   %<<<new
\def\Bbar{\ifmath{\kern 0.18em\overline{\kern -0.18em B}}{}} %<<<new
\def\Dbar{\ifmath{\kern 0.2em\overline{\kern -0.2em D}}{}}   %<<<new
\def\Y#1S{\ifmath{\Upsilon\rm(#1S)}} %<<<new

\def\kev  {\ifmath{\mbox{\,ke\kern -0.08em V}}} %<<<
\def\mev  {\ifmath{\mbox{\,Me\kern -0.08em V}}} %<<<
\def\gev  {\ifmath{\mbox{\,Ge\kern -0.08em V}}} %<<<
\def\gevc {\ifmath{\mbox{\,Ge\kern -0.08em V$\!/c$}}} %<<<
\def\mevc {\ifmath{\mbox{\,Me\kern -0.08em V$\!/c$}}} %<<<
\def\gevcc{\ifmath{\mbox{\,Ge\kern -0.08em V$\!/c^2$}}} %<<<
\def\mevcc{\ifmath{\mbox{\,Me\kern -0.08em V$\!/c^2$}}} %<<<

\begin{document}
{\small
\begin{flushright} 
        BABAR-PROC-04/012 \\
%        SLAC-PUB xxx \\
        LPNHE 2004-08 \\
        hep-ex/xxxxxxx\\[0.3cm]
\end{flushright} 
}

\vspace*{3cm}
\title{Measurements related to CKM angle $\alpha$ in \babar.}

\author{L.~Roos\\
On behalf of the \babar~ collaboration
} 

\address{Laboratoire de Physique Nucl\'eaire et de Hautes Energies,
                   IN2P3-CNRS et Universit\'es Paris VI et Paris VII,
                   4 place Jussieu, F-75252 Paris Cedex 05, FRANCE}

\maketitle\abstracts{The \babar\ collaboration measurements of the $B\to \pi\pi$,
$B \to \rho\pi$ and $B \to \rho\rho$ decays are presented. New results, from a 
$113~{\rm fb^{-1}}$ data sample, on the time-dependent $CP$ asymmetries of the
longitudinally polarized component of the $\Bz \to \rho^+\rho^-$ channel are
$S_{\rho\rho,long}=-0.19\pm0.33\pm0.11$
and $C_{\rho\rho,long}=-0.23\pm0.24\pm0.14$.
Constraints on the Unitarity Triangle angle $\alpha$ from the $\pi\pi$ and
the $\rho\rho$ systems are derived.}
%%%%%%%%%%%%%%%%%%%%%%%%%%%%%%%%%%%%%%%%%
%%%%%%%%%%%%%%%%%%%%%%%%%%%%%%%%%%%%%%%%%

The \babar~ and Belle experiments have reported in 2001
the first observation of $CP$ violation in the $B$ meson 
system~\cite{s2betaBabar,s2betaBelle}. By measuring the value of the $CP$ 
parameter $\sin2\beta$, they have provided the first direct contraint on one of the
Unitarity Triangle (UT) angles. In order to check the consistency of 
the $CP$ violation description in the Standard Model, it is of main importance 
to measure the other angles of the UT. In this paper, we describe 
the measurements by the \babar~ experiment of the time-dependent $CP$-violating
 asymmetries in three decay modes,
$B \rightarrow \pi\pi$, $B \rightarrow \rho\pi$ and $B \rightarrow \rho\rho$, 
related to the CKM angle 
$\alpha = {\rm arg}\left[ - \frac{V_{td}V_{tb}^*}{V_{ud}V_{ub}^*} \right]$.

\section{Extraction of $\alpha$ from the decays $B \rightarrow hh'$ ($h,h'=\pi,\rho$)}

\subsection{Basic Formulae}

\begin{figure}[h]
  \centerline{
        \epsfxsize3.5cm\epsffile{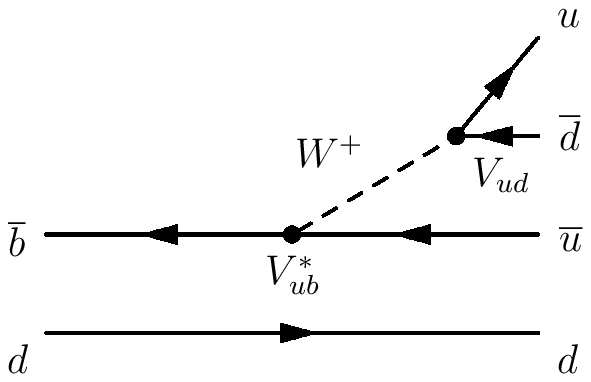}
        \hspace{2cm}
        \epsfxsize3.5cm\epsffile{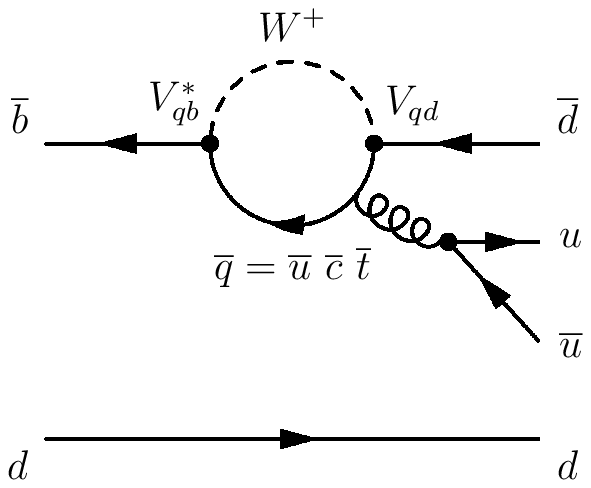}
  }
  \vspace{0.cm}
  \caption[.]{\label{fig:B0pippim}\em
        Tree (left) and penguin (right) diagrams for the decays
        $\Bz\to\pi^\pm\pi^\mp$, $\Bz\to\rho^\pm\pi^\mp$ and $\Bz\to\rho^\pm\rho^\mp$.}
\end{figure}

The decay of a neutral $B$ meson into two identical particles $B \rightarrow
hh$ ($h=\pi$ or $\rho$
\footnote{Due to the finite width of the $\rho$, the two mesons in the $B\to \rho\rho$ in
the final state are not necessarily      identical. See the comment on this approximation
in Section \ref{sec:constraints}.}
 occurs via two topologies, illustrated in 
Fig. \ref{fig:B0pippim}: a tree-level process (left) and  
one-loop penguin diagrams (right). The $CP$ parameter
$\lambda$, defined by 
$\lambda =   \frac{q}{p} 
             \frac{\bar A}{A}$,
where $q$ and $p$ are the complex coefficients that link the mass and the 
flavour eigenstates in the $B$ system, and $A$ (resp. $\bar A$) is the $\Bz$ 
(resp. $\barB^0$) decay amplitude, 
 can be expressed in terms of $\alpha$ as

\begin{equation}
\lambda = e^{2i\alpha} \frac{1-\frac{|V_{td}^*V_{tb}|}{|V^*_{ud}V_{ub}|}P/T e^{-i\alpha}}
                            {1-\frac{|V_{td}^*V_{tb}|}{|V^*_{ud}V_{ub}|}P/T e^{i\alpha}}~.
\end{equation}

$T$ and $P$ are complex amplitudes dominated respectively by the tree
and the penguin topologies. 

Experimentally, one measures the time-dependent decay rate 

\begin{equation}
\label{eq:pipiDt}
f_{Q_{tag}}(\Delta t) =
\frac{e^{-|\Delta t|/\tau}}{4\tau}
\left[ 1
+ Q_{tag} S_{hh}\sin(\Delta m_d\Delta t) 
- Q_{tag} C_{hh}\cos(\Delta m_d\Delta t)
\right],
\end{equation}

where $\Delta t$ is the decay time difference between the $B$ decaying to the 
$hh$ final state and the second $B$ in the event, denoted $B_{tag}$. $\tau$ 
is the neutral $B$ lifetime and $\Delta m_d$ is the $B^0{\barB}^0$ oscillation 
frequency.  $Q_{tag}$ is set $1$ ($-1$) if the 
$B_{tag}$ is a $B^0$ (${\bar B}^0$). The $CP$-violating asymmetries
$S_{hh}$ and $C_{hh}$ are related to the parameter $\lambda$ by

\begin{equation}
S_{hh} = \frac{2\Im m \lambda}{1+|\lambda|^2}, \hspace{1cm}
C_{hh} = \frac{1-|\lambda|^2}{1+|\lambda|^2}.
\end{equation}

$S_{hh}$ reflects the $CP$ violation induced by the interference between 
the  mixing and the decay; $C_{hh}$ is the direct 
$CP$-violating
asymmetry and comes from the interference between decay processes. In absence
of penguin contributions, $C_{hh}$ vanishes and $S_{hh}$ is simply
related to the CKM angle $\alpha$ by $S_{hh} = \sin(2\alpha)$.

In the more general case of the $\Bz(\barB^0) \rightarrow \rho^\pm\pi^\mp$ decay, the 
time-dependent decay rate reads

\beqn
\label{eq:rhopiDt}
f^{\rho^\pm \pi^\mp}_{Q_{tag}}(\Delta t) = (1 \pm A_{\rho\pi}) 
\frac{e^{-|\Delta t|/\tau}}{4\tau}
[ 1
& + & Q_{tag}( S_{\rho\pi} \pm \Delta S_{\rho\pi} )\sin(\Delta m_d\Delta t) \nonumber\\
& - & Q_{tag}( C_{\rho\pi} \pm \Delta C_{\rho\pi} )\cos(\Delta m_d\Delta t)
],\\
\nonumber
\eeqn

where the $\pm$ sign depends on whether the $\rho$ meson is emitted
by the $W$ boson or comes from the spectator quark. $A_{\rho\pi}$ is a direct 
\CP~ violation parameter, measuring the asymmetry between the $\rho^+\pi^-$
 and $\rho^-\pi^+$ final states, whereas $\Delta S_{\rho\pi}$ 
and $\Delta C_{\rho\pi}$, which arise from the fact that two  production
modes of the $\rho$ are possible, are dilution terms and have no \CP~ content.

\subsection{The Isospin Analysis}

Using strong isospin symmetry, one can extract $\alpha$ up to discrete 
ambiguities from the \CP-violating asymmetries defined above~\cite{lnqs}.
The decay amplitudes of the isospin-related final states 
obey the pentagonal relations

\beqn
\label{eq:pentagon}
    \sqrt{2}\left(\Apz+\Azp\right)      &=& 2\Azz+\Apm +\Amp~, \\
\label{eq:pentagonBar}
    \sqrt{2}\left(\Apzb+\Azpb\right)    &=& 2\Azzb+\Apmb +\Ampb~.
\eeqn

where $A^{ij}_{\rho\pi} = A(B^0$ or $B^+ \rightarrow \rho^i \pi^j)$, 
$\bar A^{ij}_{\rho\pi} = A(\bar B^0$ or $B^- \rightarrow \rho^i \pi^j)$, 
 $i,j=+,-$ or $0$. With the use of these relations, 
 12 unknows (6 complex amplitudes with one arbitrary phase, and the CKM angle $\alpha$) are 
to be determined
while 13 observables are available: $S_{\rho\pi}$, $C_{\rho\pi}$, 
$\Delta S_{\rho\pi}$, $\Delta C_{\rho\pi}$, $A_{\rho\pi}$; the average branching
fractions ${\cal B}(B^0/\barB^0 \rightarrow \rho^\pm\pi^\mp)$, 
${\cal B}(B^0/\barB^0 \rightarrow \rho^0\pi^0)$, 
${\cal B}(B^+ \rightarrow \rho^+\pi^0)$, ${\cal B}(B^+ \rightarrow \rho^0\pi^+)$;
two time-dependent \CP-violating asymmetries in the $\Bz \rightarrow \rho^0
\pi^0$ decay ($S^{00}_{\rho\pi}$, $C^{00}_{\rho\pi}$) and two direct 
\CP~ asymmetries in $B^+ \rightarrow \rho^+
\pi^0$ and $B^+ \rightarrow \rho^0\pi^+$.

In the case of two identical mesons in the final state, 
Eqs. (\ref{eq:pentagon},\ref{eq:pentagonBar}) simplify to two triangular relations
\beqn
\label{eq:triangular}
  \Apippiz      &=& \frac{1}{\sqrt{2}}\Apipi + \Apizpiz~, \\
  \Abarpippiz   &=& \frac{1}{\sqrt{2}}\Abarpipi + \Abarpizpiz~.
\eeqn

The information counting leads then to 6 unknowns and 7 observables: 
three branching fractions 
${\cal B}(\Bz \rightarrow h^+h^-)$,
${\cal B}(B^+ \rightarrow h^+h^0)$,
${\cal B}(\Bz \rightarrow h^0h^0)$;
$S_{hh}$, $C_{hh}$, $S^{00}_{hh}$, $C^{00}_{hh}$. In the $\pi\pi$ system,
$S^{00}_{hh}$ is hard or impossible to measure and one is left with 6 
observables: $\alpha$ can be extracted with an 8-fold ambiguity within $[0,\pi]$
~\cite{grolo}.

At present, $S^{00}_{hh}$ and  $C^{00}_{hh}$ have not been measured, neither 
in the $\pi\pi$ nor in the $\rho\rho$ system. Therefore, one cannot 
measure $\alpha$ but rather set a bound~\cite{glss}

\beq
\label{eq:boundPZ}
   {\rm cos}\,(2\alpha-2\alpha_{\rm eff}) \ge
        \frac{1}{D}\left(1-2\frac{\BRpizpiz}{\BRpippiz}\right)
        + \frac{1}{D}\frac{\left(\BRpipi-2\BRpippiz+2\BRpizpiz\right)^2}
                          {4\BRpipi\BRpippiz }~,
\eeq

where the effective angle $\alpha_{\rm eff}$ is defined by 
$\alpha_{\rm eff}\equiv {\rm arg}(\lambda)$ and $D\equiv\sqrt{1-C_{hh}^2}$. 
Note that Eq. (\ref{eq:boundPZ}) fully 
exploits the isospin relations while the well-known Grossman-Quinn
bound~\cite{gq} is recovered by neglecting the second term on the 
right-hand side of Eq. (\ref{eq:boundPZ}) and setting $D$ to 1.

\section{Data Analysis}

\subsection{Data Selection}

Results on $B \rightarrow \pi\pi$, $B \rightarrow \rho\pi$ and 
$B \rightarrow \rho\rho$ decays are presented. Signal events are selected
by combining the relevant number of charged tracks and/or neutral clusters 
to form a $B$ candidate. Other particles in the event form the $B_{tag}$. A
vertexing algorithm~\cite{s2btaPRD} is used to determine  the decay time 
difference  $\Delta t$ between the two $B's$ from their distance along the
$z$ axis ($\Delta z$). The typical resolution on
$\Delta z$ is $180~\mu$m.
The tagging procedure, based on a multivariate technique~\cite{tagging}, 
is applied on the $B_{tag}$ to determine
the flavour of the $B$ at $\Delta t=0$. The total effective tagging efficiency
is $(28.4\pm0.7)\%$.
The data selection relies on several common aspects, which are summarized below.
 
Useful variables to discriminate signal from background are primarily:
the beam-energy-substituted mass 
$m_{ES} = \sqrt{(s/2+{\vec p_i}.{\vec p_B})^2/E_i^2-
{\vec p_B}^2}$, where $\sqrt{s}$ is the total energy in the $e^+e^-$ 
center of mass (CM), 
$(E_i,{\vec p_i})$ is the four-momentum of the initial state and 
${\vec p_B}$ the momentum of the $B$, both measured in the laboratory 
frame; the energy difference, $\Delta E$, between the CM energy of the $B$ and 
$\sqrt{s}/2$.

The topogical properties of  $B\barB$ decays in the $\Upsilon(4S)$
rest frame are used to discriminate the signal from the $B \rightarrow
q\bar q~(q=u,d,s,c)$ background.  All analyses use the $L_0$ and
$L_2$ moments,   defined
in the $\Upsilon(4S)$ rest frame as

\beq
L_0 \equiv \sum_{i\not\in B} p_i^*~,\:
L_2 \equiv \sum_{i\not\in B} p_i^* \cos^2\theta_i
\eeq

where $p_i^*$ is the momentum of particle $i$ not included in the $B$ candidate, 
and $\theta_i$ is the angle between $p_i^*$ and the thrust of the 
$B$ candidate. 
$L_0$ and $L_2$ are combined with a Fisher algorithm. In order to increase 
the discrimination, they can be further combined with variables such as the 
cosine of the angle between the beam axis and the $B$ candidate momentum or the
$B$ thrust axis.

All channels but $B^0 \rightarrow \pi^+\pi^-$ decays suffer from $B$ background.
 In addition to $\Delta E$, other discriminating variables are  the $\rho$ 
candidate mass 
($0.4<m(\pi^+\pi^0)<1.3~{\rm GeV/c^2}$ or $0.53<m(\pi^+\pi^0)<0.9~{\rm GeV/c^2}$)
and the helicity angle ($|cos\theta_\rho|>0.25$, where $\theta_\rho$ is the angle of
one daughter pion mementum and the $B$ momentum in the $\rho$ rest frame).

Particle identification mainly relies on the DIRC~\cite{babar}, the Cherenkov 
detector, which
provides a kaon-pion separation greater than $2.1\sigma$ over a $[1.7-4.2]$ GeV/c 
momentum range.

An unbinned likelihood fit is finally performed on selected events: 
for each event, a probability density function is built from discriminating 
variables, including the $\Delta t$-dependence, either in its simple exponential
form for charged $B's$ or following Eqs. (\ref{eq:pipiDt}) and (\ref{eq:rhopiDt}). 

\subsection{Results}

Branching fraction and time-dependent $CP$ asymmetries of the 
$B \rightarrow \pi\pi$, $B \rightarrow \rho\pi$ and $B \rightarrow \rho\rho$ 
decays are summarized in Table \ref{tab:all}. 

The branching fractions of the three isospin partner decays in the $\pi\pi$ system
are measured~\cite{brpippim,brpippi0,brpi0pi0}. The statistical significance of 
the recently observed $\Bz
\to \pi^0\pi^0$ mode is $4.2\sigma$ ~\cite{brpi0pi0}. No evidence for $CP$ violation is observed
in the $\Bz\to \pi^+\pi^-$ channel~\cite{cppippim}.

The quasi-two-body analysis parameters of the $\Bz\to \rho^\pm\pi^\mp$ decay are also 
reported~\cite{rhopi}. Direct $CP$ violation information carried by
the $C_{\rho\pi}$ and $A_{\rho\pi}$ parameters has a $2.5\sigma$ significance,
which is likely to be a statistical fluctuation since the well measured 
branching fraction of the $\Bz\to \rho^+K^-$ decay is rather small 
(${\cal B}(\Bz\to \rho^+K^-) = (9.0\pm1.6)~10^{-6}$)~\cite{rhopi}.

The $\Bz \to \rho^+\rho^-$ and $B^+ \to \rho^+\rho^0$ modes are observed~\cite{prlRhorho,gritsan},
while only an upper limit is set on the branching fraction of the
$\Bz \to \rho^0\rho^0$ channel~\cite{gritsan}.
Dominance of the longitudinally polarized component in the first two decays 
is observed. Recently, the \babar\ 
collaboration has reported on the measurement of the 
$CP$ violating asymmetries in the $\Bz\to \rho^+\rho^-$ longitudinal component
decay on $81~{\rm fb^{-1}}$ ~\cite{prlRhorho,gregory}. The measurement
has been updated on a $113~{\rm fb^{-1}}$ sample and found in agreement with the
previous result. A detailed analysis of the background due to other $B$ decays
is performed. The main systematic uncertainty on the asymmetries $S_{\rho\rho,long}$
and $C_{\rho\rho,long}$ is found to be the unknown $CP$ violation in $B$ background
events.

\begin{table}[t]
\begin{center}
\setlength{\tabcolsep}{0.0pc}
{\normalsize
\begin{tabular*}{\textwidth}{@{\extracolsep{\fill}}ccc}\hline
&& \\[-0.4cm]
 ${\cal B}(B^0 \rightarrow \pi^+\pi^-)(^*)$  &  ${\cal B}(B^+ \rightarrow \pi^+\pi^0)(^*)$ 
                           &    ${\cal B}(B^0 \rightarrow \pi^0\pi^0)~$ \\[0.1cm]
&& \\[-0.4cm]
 $4.7\pm0.6\pm0.2$   & $5.5^{\,+1.0}_{\,-0.9}{\pm0.6}$ &  $2.1\pm0.6\pm0.3$  \\[0.1cm]
\hline
&& \\[-0.4cm]
$C_{\pi\pi}$  &    $S_{\pi\pi}$  &   Correlation coeff.  \\[0.1cm]
 $-0.19\pm0.19\pm0.05$    & $-0.40\pm0.22\pm0.03$  & $-0.02 $    \\[0.15cm]
\hline 
\end{tabular*}
\begin{tabular*}{\textwidth}{@{\extracolsep{\fill}}ccc}\hline
&& \\[-0.4cm]
$C_{\rho\pi}$ & $S_{\rho\pi}$  &   $A_{\rho\pi}$ \\[0.1cm]
 $ 0.35\pm0.13\pm0.05$    & $-0.13\pm0.18\pm0.04$  & $-0.114\pm0.062\pm0.027$ \\[0.15cm]
\hline 
\end{tabular*}
\begin{tabular*}{\textwidth}{@{\extracolsep{\fill}}ccccc}
&&&& \\[-0.4cm]
& $\Delta C_{\rho\pi}$ & & $\Delta S_{\rho\pi}$ & \\[0.1cm]
& $ 0.20\pm0.13\pm0.05$    & & $ 0.33\pm0.18\pm0.03$  &\\[0.15cm]
\hline
\end{tabular*}
\begin{tabular*}{\textwidth}{@{\extracolsep{\fill}}cccc}\hline
&&& \\[-0.4cm]
 $C_{\rho\rho,long}$ & $S_{\rho\rho,long}$ & 
     ${\cal B}(B^0 \rightarrow \rho^+\rho^-)~(^*)$ & $f_L(B^0 \rightarrow \rho^+\rho^-)~(^*)$ \\[0.1cm]
 $-0.23\pm0.24\pm0.14$ & $-0.19\pm0.33\pm0.11$  &
     $30\pm4\pm5$ & $0.99\pm0.03^{+0.04}_{-0.03}$\\[0.15cm]
\hline
\end{tabular*}
\begin{tabular*}{\textwidth}{@{\extracolsep{\fill}}ccc}
&& \\[-0.4cm]
 ${\cal B}(B^+ \rightarrow \rho^+\rho^0)~(^*)$ & $f_L(B^+ \rightarrow \rho^+\rho^0)~(^*)$ &
     ${\cal B}(B^0 \rightarrow \rho^0\rho^0)~(^*)$\\[0.1cm]
 $22.5^{+5.74}_{-5.4 }\pm5.8  $ & $0.97^{+0.03}_{-0.07}\pm0.04$ &
     $<2.1~(90\%~{\rm CL})$ \\[0.15cm]
\hline
\end{tabular*}
}
\vspace{-0.4cm}
\caption[.]{\label{tab:all} \em
  Branching fractions and time-dependent $CP$ asymmetries in $B \rightarrow \pi\pi$,
  $B \rightarrow \rho\pi$ and $B \rightarrow \rho\rho$ decays.
   Measurements are performed on samples
of $81~{\rm fb^{-1}}$ (marked with $(^*)$) or $113~{\rm fb^{-1}}$. The first error is
statistical and the second is systematic. Branching fractions are given in $10^{-6}$ units.
 }
\end{center}
\end{table}

\section{Constraints on $\alpha$}
\label{sec:constraints}
At present, SU(2)-based analysis  of the $B \to \rho\pi$ system does not lead  to 
useful constraint on $\alpha$, since the construction of the pentagons described by
Eqs. (\ref{eq:pentagon}) requires more precise measurements than currently available. 
Data sample with a 
luminosity of the order of
$10~{\rm fb^{-1}}$ is needed. More promising is a Dalitz plot analysis that would bring
informations on the strong phases involved in the $\Bz \to \rho^\pm\pi^\mp$ decay.
If the validity of QCD Factorization was established, such a model could also help
to constraint $\alpha$, with an accuracy of $9^\circ$ with current data~\cite{bn}.

\begin{figure}[t]
  \centerline{
        \epsfxsize8.cm\epsffile{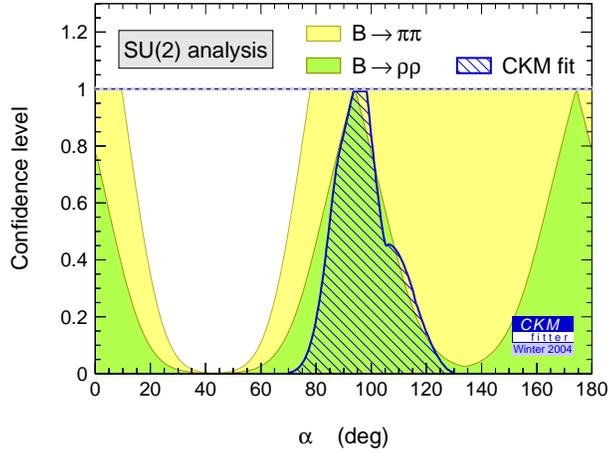}
  }
  \vspace{-0.5cm}
  \caption[.]{\label{theOnlyFig}\em
Confidence level from the SU(2) analysis of the $B\to\pi\pi$ (light shaded)
and $B\to\rho\rho$ (dark shaded) decays as a function of $\alpha$. Also
shown is the result from the standard CKM fit (hatched area, see text).}
\end{figure}

The confidence level as a function of $\alpha$ obtained from the isospin analysis of
the $B \to \pi\pi$ decays is shown on Fig. \ref{theOnlyFig} (light shaded 
histogram)~\cite{thePap}. The $CP$ asymmetries $C_{\pi\pi}$ and $S_{\pi\pi}$ quoted
in Table \ref{tab:all} are used, together with the world average values of the
branching fractions of the
$\Bz \to \pi^+\pi^-$, $B^+ \to \pi^+\pi^0$ and $\Bz \to \pi^0\pi^0$ channels~\cite{hfag}.
The plateau reflects the unfruitful bound on $\alpha-\alpha_{\rm eff}$:
$-54^\circ<\alpha-\alpha_{\rm eff}<52^\circ$ ($90\%$ CL), largely dominated by the 
uncertainty on the penguin contribution.\footnote{
A strict application of Eq. (\ref{eq:boundPZ}) gives  a symmetric bound on  
$\alpha-\alpha_{\rm eff}$. However, in the study of Ref. \cite{thePap}, electroweak
penguins are taken into account, following the recipe proposed by Neubert and Rosner
\cite{nrpew} (so that no additional degrees of freedom is introduced), leading to the
asymmetric bound given above.
}

Similarly, one can apply the isospin analysis to the longitudinal components of the
$B\to\rho\rho$ decays. \babar\ measurements of the time-dependent asymmetries, branching
fraction and polarization fraction
in $\Bz \to \rho^+\rho^-$ mode (Tab. \ref{tab:all}) are used, as well as the \babar\
and Belle average branching fraction~\cite{hfag} and polarization 
fraction~\cite{gritsan,belleRhorho} for the $B^+ \to \rho^+\rho^0$ channel. The analysis
includes the
value leading to the upper limit on ${\cal B}(\Bz \to \rho^0\rho^0)$ quoted in
Tab. \ref{tab:all}, ${\cal B}(\Bz \to \rho^0\rho^0)=(0.6^{+0.7}_{-0.6}\pm0.1)~10^{-6}$,
and it is assumed conservatively that the decay is $100\%$ longitudinally polarized.
Interference with higher radial excitations of the $\rho$ meson, non-resonant contributions
or possible isospin violating effects due to the finite width of the $\rho$ are 
neglected~\cite{ligetiquinn}. Powerful constraint on $\alpha$ is obtained
 (Fig. \ref{theOnlyFig}, dark shaded
histogram), in agreement with and with comparable accuracy to the standard CKM fit, which includes
the constraints from $CP$ violation measurements in neutral kaon mixing,
$|V_{ub}|$, $|V_{cb}|$, $\Bz\Bzb$ and $\Bs\Bsb$ oscillations, and $\sin2\beta$ (hatched 
area)~\cite{thePap}. Choosing the solution closest to the standard fit constraint, one estimates
$\alpha=(96\pm10_{stat}\pm4_{syst}\pm13_{peng})^\circ$. Note that the peak-like shape of the CL
function, in contrast with the plateau expected from Eq. (\ref{eq:boundPZ}), is due central values
violating the isospin relations (\ref{eq:triangular}). However, this ``incompatibility'' is 
well covered by the present experimental uncertainties.

\section{Conclusion}

The \babar\ collaboration has published evidence or observation of the three decay modes of the 
$B \to \pi\pi$ system. The measurements
of the time-dependent $CP$ asymmetries in $\Bz \to \pi^+\pi^-$ channel do not lead to a useful
constraint on $\alpha$, due to the present uncertainty on the penguin contribution.
The quasi-two-body $CP$ asymmetries in the $B \to \rho\pi$ decay have been measured. Next
step is to perform the Dalitz plot analysis  in order to constrain the
strong phases involved in the decay and that are needed to extract $\alpha$. In contrast with
 the $\pi\pi$ and $\rho\pi$ systems, a powerful constraint on $\alpha$  in obtained from 
the measurements of the time-dependent asymmetries of the longitudinally polarized component
of the $\Bz \to \rho^+\rho^-$ channel. Performing the isospin analysis and choosing
the solution closest to the standard CKM fit, \babar~ quotes 
$\alpha=(96\pm10_{stat}\pm4_{syst}\pm13_{peng})^\circ$.

%%%%%%%%%%%%%%%%%%%%%%%%%%%%% REFERENCES %%%%%%%%%%%%%%%%%%%%%%%%%%%%%%


\begin{thebibliography}{99}

\bibitem{s2betaBabar}
       \babar~ collaboration, \PRL\   {\bf 87} 091801 (2001).
\bibitem{s2betaBelle}
        Belle~ collaboration, \PRL\   {\bf    87} 091802 (2001).
\bibitem{lnqs}          
        H.J.~Lipkin, Y.~Nir, H.R.~Quinn and A.~Snyder,
                        \PRD\      {\bf 44} 1454 (1991)

\bibitem{grolo}
        M.~Gronau and D.~London, 
                        \PRL\ {\bf 65} 3381 (1990)

\bibitem{glss}
        M.~Gronau, D.~London, N.~Sinha and R.~Sinha,
                        \PLB\         {\bf 514} 315 (2001)

\bibitem{gq}            
        Y.~Grossman and H.R.~Quinn, 
                        \PRD\ {\bf 58}, 017504 (1998) 

\bibitem{s2btaPRD}      
        \babar\ collaboration (B.~Aubert et al.)
                        \PRD\ {\bf 66}, 032003   (2002) 

\bibitem{tagging} 
         \babar\ collaboration (B.~Aubert et al.),
                        \PRL\ {\bf 89} 201802 (2002)

\bibitem{babar}
         \babar\ collaboration (B.~Aubert et al.),
                        \NIM\ {\bf A}479, 1 (2002).

\bibitem{brpippim}
         \babar\ collaboration (B.~Aubert et al.),
                        \PRL\ {\bf 89} 281802 (2002)

\bibitem{brpippi0}
         \babar\ collaboration (B.~Aubert et al.),
                        \PRL\ {\bf 91} 021801 (2003)

\bibitem{brpi0pi0}
         \babar\ collaboration (B.~Aubert et al.),
                        \PRL\ {\bf 91} 241801 (2003)

\bibitem{cppippim}
         H.~Jawahery, 
                        Talk given at Lepton-Photon 2003, Batavia, Illinois, 11-16 Aug 2003

\bibitem{rhopi}
         \babar\   collaboration (B.~Aubert et al.), 
                        Phys. Rev. Lett. {\bf 91}, 201802 (2003);
                        updated results at http://www.slac.stanford.edu/BFROOT/www/doc,
                        \babar-PLOT-0055 (2003)

\bibitem{prlRhorho}
         \babar\ collaboration (B.~Aubert et al.), 
                        \babar-PUB-04-09, hep-ex/0404029 (2004)

\bibitem{gritsan}
         \babar\  collaboration (B. Aubert et al.),
                        \PRL\  {\bf 91}, 171802 (2003)

\bibitem{gregory}
         G. Schott (for the \babar\ collaboration), talk given at Les Rencontres
         de physique de la Vall\'ee d'Aoste, La Thuile, Italy, 29 Feb - 6 March 2004 

\bibitem{bn}
         M.~Beneke and M.~Neubert,
         {\em Nucl. Phys.} {\bf B675} 333 (2003)

\bibitem{thePap}
         The CKMfitter Group (J. Charles et al.),
         CPT-2004/P.030, LAL 04-21, LAPP-EX-2004-01, 
         LPNHE 2004-01, hep-ph/0406184

\bibitem{nrpew}
         M.~Neubert, J.L.~Rosner, 
                        \PLB\  {\bf 441}, 403 (1998); 
                        \PRL\  {\bf 81}, 5076 (1998)

\bibitem{hfag}
          The Heavy Flavor Averaging Group,
                        http://www.slac.stanford.edu/xorg/hfag/

\bibitem{belleRhorho}
        Belle~ collaboration (J. Zhang, M. Nakao et al.),
        \PRL\   {\bf 91} 221801 (2003)

\bibitem{ligetiquinn}   Y.~Grossman, Z.~Ligeti, Y.~Nir and H.~Quinn
                        Phys. Rev. {\bf D68}, 015004 (2003)

\end{thebibliography}
\end{document}
%%%%%%%%%%%%%%%%%%%%%%%%%%%%% END OF DOCUMENT %%%%%%%%%%%%%%%%%%%%%%%%%